\nonstopmode

\newcommand{\XeFn}{XeF$_n$}
\newcommand{\XeFt}{XeF$_2$}
\newcommand{\XeFf}{XeF$_4$}
\newcommand{\XeFs}{XeF$_6$}
\newcommand{\Ft}{F$_2$}
\newcommand{\U}[1]{\,{\rm #1}}
\newcommand{\I}[1]{_{\rm #1}}

\newcommand{\ie}{i.e.{}}

\newcommand{\etal}{\textit{et al.{}}}
\newcommand{\angstrom}{\U{\hbox{\AA}}}
\newcommand{\degree}{{}^{\circ}}

\documentclass[prb,letterpaper,twocolumn]{revtex4}
\usepackage{graphicx}

\begin{document}
\title{Ionization of the Xenon Fluorides}
\date{16 June 2003}
\author{Christian Buth}
\altaffiliation[Author to whom all correspondence should be addressed.
Present address: ]{Max-Planck-Institut f\"ur
Physik komplexer Systeme, N\"othnitzer Stra\ss{}e~38,
01187~Dresden, Germany}
\email[electronic mail: ]{Christian.Buth@web.de}
\author{Robin Santra}
\altaffiliation[Present address: ]{JILA, University of Colorado, Boulder, CO
80309-0440, USA}
\author{Lorenz S. Cederbaum}
\affiliation{Theoretische Chemie, Physikalisch-Chemisches
Institut, Ruprecht-Karls-Universit\"at Heidelberg, Im Neuenheimer
Feld~229, 69120~Heidelberg, Germany}

\begin{abstract}
The noble-gas atom xenon can bind fluorine atoms and form several stable
compounds. We study the electronic structure of the xenon fluorides
(XeF$_n$, $n = 2,4,6$) by calculating their ionization spectra using a
Green's function method, which allows one to treat many-body effects at a
high level. Our focus is on the valence region and on the Xe$\,4d$~core
hole. We observe a sensitive dependence of the spectra on the number of
fluorine ligands. Systematic line shifts are uncovered and explained.
In the Xe$\,5s$ and F$\,2s$~inner-valence regimes, from XeF$_2$ to
XeF$_6$, the usefulness of the one-particle picture of ionization is
found to become progressively worse. Moreover, adding the electronegative
fluorine ligands seems to enhance---not suppress---the Auger decay of a
Xe$\,4d$~core hole.
\end{abstract}
\maketitle

\section{Introduction}

Noble-gas atoms are, in general, chemically rather inert.
Nevertheless, since the heavy noble-gas species krypton, xenon,
and radon possess a comparatively small ionization potential, they
\emph{can} form molecules, at least together with electronegative
elements. The most stable and most widely investigated among these
compounds are the xenon fluorides~\cite{Holleman:IC-01} XeF$_2$,
XeF$_4$, and XeF$_6$.

The electronic structure of the xenon fluorides has been the subject of a
number of both experimental and theoretical investigations. In the 1970's,
photoabsorption experiments, at photon energies between $50$ and $160$~eV
\cite{Comes:XeF-73,Nielsen:XeF-74} and between $6$ and $35$~eV
\cite{Nielsen:XeF-76}, were carried out at DESY in Hamburg. 
Carroll~\etal~\cite{Carroll:XeF-74} employed photoelectron spectroscopy to
measure the chemical shifts of the F$\,1s$ and Xe$\,3d$ levels in the xenon
fluorides. More recently, using synchrotron radiation, 
Cutler~\etal~\cite{Cutler:Xe-91} obtained gas-phase photoelectron spectra 
with such a high resolution that fluorine-ligand field splittings on the 
Xe$\,4d$ levels could be extracted.

Photoelectron spectroscopy, in combination with \emph{ab initio} studies,
is an outstanding tool for characterizing the electronic structure of a
molecule. However, at least in the case of the xenon fluorides, there
exist---to our knowledge---only few theoretical contributions to motivate
and back-up experimental work. The pioneering
papers~\cite{Basch:XeF-71,Scheire:XeF-80,Gutsev:XeF-81}, which are more
than 20 years old, utilized self-consistent-field methods to calculate
ionization potentials. The restriction to an effective one-particle model
made it impossible to uncover complexities due to electron correlation.
Newer theoretical work~\cite{Kaupp:Xe-96,DFB:XeF-97,Liao:XeF-98} on xenon
fluorides focused on electronic ground-state geometries and dissociation
energies. It is therefore timely to improve on the early studies and to
explore some of the \emph{many-body} physics of electrons in
XeF$_2$, XeF$_4$, and XeF$_6$. This is the purpose of this paper.

We proceed as follows. Section~\ref{sec:computation} describes the
Green's function method we use to calculate ionization spectra.
We discuss the molecular geometries we employed in our calculations,
and we analyze the impact of relativistic effects on the ionization
spectra of the xenon fluorides. The ionization spectra we calculated
cover the valence regime and the Xe$\,4d$~core line; they are
presented in Sec.~\ref{sec:results}. A detailed comparison between
spectra that take electron correlation into account and spectra based on
the Hartree-Fock model demonstrates the occurrence of many-body phenomena
in almost the entire spectral regime we consider. These phenomena
display an interesting dependence on the number of fluorine ligands.
A summary and conclusions are given in Sec.~\ref{sec:conclusion}.

\section{Computational Details}
\label{sec:computation}
\subsection{Algebraic Diagrammatic Construction}
\label{sec:pADC}

Green's functions are a fundamental tool of many-body
theory.~\cite{Fetter:MP-71,Mattuck:FD-76,Gross:MP-91,Szabo:MQC-82}
They are well suited to calculating various properties of
molecules.~\cite{Cederbaum:TA-77,Cederbaum:GF-98} For instance, the
pole positions of the one-particle propagator yield the vertical
ionization potentials of a molecule. The residuum or pole strength
of a pole of the one-particle propagator is a measure of the one-particle
character of a specific ionized state, \ie, it is a measure how well the
physical state is described within the one-hole configuration space
deriving from the Hartree-Fock ground state.

Algebraic diagrammatic construction~(ADC) is a sophisticated, systematic
approximation scheme for Green's functions.~\cite{Schirmer:PP-82,
Schirmer:GF-83,Schirmer:SE-89} The $n$-th order ADC~scheme, ADC($n$) for
short, contains infinite summations of those classes of Feynman diagrams
that derive from the first~$n$ orders of the Feynman-Dyson perturbation
series. The problem of finding the pole positions and pole strengths
of a Green's function is formulated in terms of the solution of a Hermitian
matrix eigenvalue problem.

The one-particle ADC(3)~scheme we use in this work to calculate ionization
spectra employs the Dyson equation,~\cite{Fetter:MP-71,Schirmer:GF-83}
which allows one to sum many diagram classes in addition to the summation
already carried out by the ADC scheme itself. Exploiting the validity of the
Dyson equation necessitates the combined treatment of the ionization
potentials \emph{and} the electron affinities. The latter are also given by
pole positions of the one-particle propagator. This approach enlarges the
aforementioned ADC eigenvalue problem considerably. Therefore, the affinity
block in the ADC matrix is reduced in practice by performing on this block
a few (typically ten) block Lanczos iterations.~\cite{Weikert:BL-96} The
most important spectral features of the affinity block are preserved in
this way, and the final eigenvalue problem to be solved becomes much
more manageable.

The ionization potentials associated with the Xe$\, 4d$~orbitals in the
xenon fluorides lie far above the molecular double ionization threshold.
Hence, not only two-hole--one-particle configurations but also
three-hole--two-particle configurations are expected to have impact on
the description of the core-level ionization spectra.~\cite{Cederbaum:CH-80,
Angonoa:KS-87,Schirmer:ISR-01} The ADC(3)~scheme does not
contain three-hole--two-particle configurations explicitly, but the
next-order scheme, ADC(4), does. Thus, one expects that the ADC(4)~scheme
yields an appreciable improvement of calculated core-level ionization
spectra.

The inclusion of three-hole--two-particle configurations enlarges the
configuration space---and the associated eigenvalue problem---extremely.
The available computer resources would not be sufficient to calculate
ionization spectra of molecules like~\XeFs\ in an acceptable amount of time.
The existing ADC(4)~programs~\cite{Tarantelli:BP-03,Angonoa:KS-87,
Schirmer:ISR-01} therefore utilize the core--valence separation
approximation~\cite{Cederbaum:CH-80} for core-level ionization, which reduces
the size of the configuration space. This approximation, however, makes it
impossible to calculate ionization potentials in the valence regime.
Furthermore, since core--valence separation implies that all elements
of the configuration space carry a core hole, those two-hole--one-particle
configurations that are needed to describe core-hole decay are not included.
For these reasons, the ADC(4)~programs are useless for the questions we would
like to address here. Hence, all ionization spectra were calculated within the
ADC(3)~scheme.~\cite{Schirmer:GF-83,Schirmer:SE-89}

\subsection{Geometries and Basis Sets}

The ground-state geometries of~\Ft, \XeFt, \XeFf, and \XeFs{} are
taken from the literature. The distance of the fluorine atoms in the
fluorine molecule is~$r($F--F$) = 1.42 \angstrom
\,$.~\cite{Holleman:IC-01} Xenon(II)-fluoride is a linear
molecule of $D_{\infty h}$~symmetry with a Xe--F~distance
of~$1.977 \angstrom \,$.~\cite{Holleman:IC-01} Xenon(IV)-fluoride
is square-planar ($D_{4h}$), the Xe--F~distance is~$1.94
\angstrom$.~\cite{Holleman:IC-01} All distances given refer to molecules in the gas phase.

The ground-state geometry of xenon(VI)-fluoride in the gas phase
is a distorted octahedron.~\cite{Kaupp:Xe-96,Cutler:Xe-91} It can
be described in $C_{3v}$~symmetry. Nevertheless, some of our
computations on~\XeFs\ were performed in $O_h$~symmetry to
elucidate the impact of the reduced symmetry. The notation of
Fig.~1a in Ref.~\onlinecite{Kaupp:Xe-96} is used here. Our
calculations on~\XeFs\ in~$C_{3v}$~symmetry use the bond
lengths~$r($Xe--F1$) = 1.856 \angstrom$ and $r($Xe--F4$) = 1.972
\angstrom$ and bond angles~$\alpha = 80.8\degree$ and $\beta =
112.8\degree$. These values, which were determined using MP2 and an
all-electron basis with $f$ functions, were taken from Tables~2 and 3 in
Ref.~\onlinecite{Kaupp:Xe-96}. Our computations on~\XeFs\ in $O_h$~symmetry
were performed at the bond length~$r($Xe--F$) = 1.952 \angstrom$, which is
taken from Table~1 in Ref.~\onlinecite{Kaupp:Xe-96}.

The ADC~programs~\cite{Weikert:BL-96,Tarantelli:BP-03}
(see Sec.~\ref{sec:pADC}) require molecular integrals and orbital energies.
These were obtained in Hartree-Fock calculations using the
{\sc Gamess-UK}~\cite{gamess-uk} program package. The employed software
does not exploit symmetry for the xenon atom. $D_{2h}$~symmetry is used
for~F$_2$, \XeFt, and \XeFf. The calculation of the ionization potentials
of~\XeFs\ in the ground-state geometry of $C_{3v}$~symmetry is performed in
$C_s$~symmetry; for the geometry of $O_h$~symmetry, {\sc Gamess-UK}
assumes $D_{2h}$~symmetry. Comparing the orbital energies from both
the $C_{3v}$ and the $O_h$ geometry shows that the overall positions
of the xenon-like orbitals are in good agreement in both symmetries,
but the splitting of the fluorine-like orbitals is enlarged due to the
increased interaction in $C_{3v}$~symmetry.~\cite{Buth:NH-02}

The xenon and the fluorine atoms are represented by the DZVP (DFT
orbital)~\cite{Godbout:GT-92,basislib} basis set. The quality of
the basis set can be estimated from Table~\ref{tab:Xe-orbitals}
by comparing the numerically exact Hartree-Fock orbital energies
(see the ensuing Sec.~\ref{sec:relativistic}) of xenon with those obtained
using the DZVP (DFT orbital) basis set. The shift of the orbitals, due to the
approximation introduced by the finite basis set, is~$\Delta \varepsilon\I{BS}
:= \varepsilon\I{HF,\ numeric} - \varepsilon\I{HF,\ DZVP}$.  The
shift is~$\approx 0.13 \U{eV}$ for the Xe$\, 4d$~orbitals and
even less for the valence orbitals because basis sets are usually
optimized with respect to the latter orbitals. This shift is neglected in
the following, for other sources of inaccuracies are more significant.

\subsection{Relativistic Effects}
\label{sec:relativistic}

There are three main relativistic effects one has to take into
account when examining heavy atoms like
xenon:~\cite{Balasubramanian:RE-97,Pyykko:RE-88}
\begin{enumerate}
  \item the relativistic radial contraction and energetic
    stabilization of the~$s$ and $p$~shells,
  \item the spin-orbit splitting,
  \item the relativistic radial expansion and the energetic
    destabilization of the (outer) $d$~and~$f$ shells.
\end{enumerate}
Effects~(1) and (3) are termed scalar relativistic
effects.~\cite{Kaupp:Xe-96}

As the theory of Sec.~\ref{sec:pADC} for the calculation of
ionization potentials is strictly nonrelativistic, one has to take
into account relativistic effects by a ``rule of thumb''. This is done
by performing for the xenon atom Hartree-Fock calculations with
\textsc{mchf84}~\cite{Froese:HF-78} and Dirac-Fock calculations, the
relativistic counterpart to
Hartree-Focki,~\cite{Balasubramanian:RE-97,Pyykko:RE-88} with 
\textsc{grasp92}.~\cite{Dyall:GR-89,Parpia:GR-96} Due to the spherical 
symmetry of atoms, the equations can be solved without fixed basis 
sets to arbitrary precision. This gives exact sets of relativistic and
nonrelativistic orbitals in the mean-field approximation.

In Table~\ref{tab:Xe-orbitals}, the orbital energies determined in this
way are listed together with Hartree-Fock orbital energies obtained by a
calculation employing the DZVP (DFT Orbital) basis set. By comparing the
orbital energies of the two numerical solutions of the Hartree-Fock and
Dirac-Fock equations, one can determine the size of relativistic effects
and correct for them in nonrelativistic computations of the ionization
potentials of the xenon fluorides.

In order to carry out this comparison, one has to note that the total
angular momentum~$j$ results from coupling the orbital angular momentum~$l$
with the electron spin:~$j = l \pm \frac{1}{2}$. A Dirac-Fock calculation
yields, for~$l \geq 1$, two orbitals,~\cite{Balasubramanian:RE-97} one
for~$j_+ = l + \frac{1}{2}$ and one for~$j_- = l - \frac{1}{2}$. The
$j_-$~orbital has a lower orbital energy than the $j_+$~orbital.

\renewcommand{\arraystretch}{1.2}%
\begin{table*}
  \centering
  \newcommand{\rb}[1]{\raisebox{1.5ex}[-1.5ex]{#1}}
  \begin{tabular}{c|l|l||c|l|l}
    HF orbital & \hfil $\varepsilon\I{HF,\ DZVP}$
    & \hfil $\varepsilon\I{HF,\ numeric}$ & DF orbital &
    \hfil $\varepsilon\I{DF}$ & \hfil $\bar\varepsilon
    \I{DF}$ \\
    \hline\hline
		     &		      &		       & $4d_{3/2}$
    & -2.71133	&		 \\
    \rb{$4d$} & \rb{-2.78280}  & \rb{-2.77788}  & $4d_{5/2}$
    & -2.63376	& \rb{-2.66479}	 \\
    \hline
    $5s$	     &	   -0.946253  &	    -0.944414  & $5s$ \quad\
    & -1.01014	&     -1.01014	 \\
    \hline
		     &		      &		       & $5p_{1/2}$
    & -0.492572 &		 \\
    \rb{$5p$} & \rb{-0.457894} & \rb{-0.457290} & $5p_{3/2}$
    & -0.439805 & \rb{-0.457394} \\
  \end{tabular}
  \caption{Hartree-Fock~(HF) and Dirac-Fock~(DF) orbital energies of xenon.
	   Hartree-Fock orbital energies are given for the DZVP (DFT
	   orbital)~\cite{Godbout:GT-92,basislib} basis
	   set and for the numerically exact solution.~\cite{Froese:HF-78}
	   The Dirac-Fock orbital energies also are numerically
	   exact.~\cite{Dyall:GR-89,Parpia:GR-96} $\bar \varepsilon{\I{DF}}$
	   is calculated using Eq.~(\ref{DFmean}).
	   All data are given in Hartree.}
\label{tab:Xe-orbitals}
\end{table*}
\renewcommand{\arraystretch}{1}%

The spin-orbit splitting between the $j_+$~orbital and the $j_-$~orbital
can be effectively removed by calculating a weighted mean
\begin{equation}
\label{DFmean}
  \bar \varepsilon{\I{DF}} = \frac{(2j_+ +1) \, \varepsilon\I{DF, +} 
  + (2j_- +1) \, \varepsilon\I{DF, -}}{2j_+ +1 + 2j_- + 1}
\end{equation}
of the two Dirac-Fock orbital energies.~\cite{Balasubramanian:RE-97}
This procedure facilitates comparison between the Dirac-Fock and the
Hartree-Fock calculation. The scalar relativistic shift of the
nonrelativistic orbitals is~$\Delta \varepsilon := \bar
\varepsilon\I{DF} - \varepsilon\I{HF}$. The shifts are~$3.077
\U{eV}$ for the Xe$\, 4d$, $-1.789 \U{eV}$ for the Xe$\,5s$,
and~$-0.00283 \U{eV}$ for the Xe$\,5p$ orbitals.  The sign of the
shift~$\Delta \varepsilon$ is predicted by the above-given 
rules~(1) and (3), because the energetic stabilization of~$s$ and
$p$~orbitals leads to a lowering of the Xe$\,5s$ and Xe$\,5p$
orbital energies and thus~$\Delta \varepsilon < 0$. Similarly,
$\Delta \varepsilon > 0$~holds for the shift of the Xe$\,4d$
orbital energies due to the energetic destabilization of the
(outer) $d$~orbitals.

Koopmans' theorem~\cite{Szabo:MQC-82,Koopmans:-33} states that the
ionization potentials of an atom or a molecule are approximately
given by the negative of the orbital energies, IP$\approx
-\varepsilon\I{HF}$. As the relativistic shift~$\Delta
\varepsilon$ represents a relativistic correction to the
Hartree-Fock orbital energies~$\varepsilon\I{HF,corr} =
\varepsilon\I{HF} + \Delta \varepsilon$, the ionization potentials are
corrected as follows: IP$\I{corr} = {}$IP$\I{HF} - \Delta \varepsilon$.

\begin{figure*}
  \begin{center}
    \includegraphics[width=\hsize,clip]{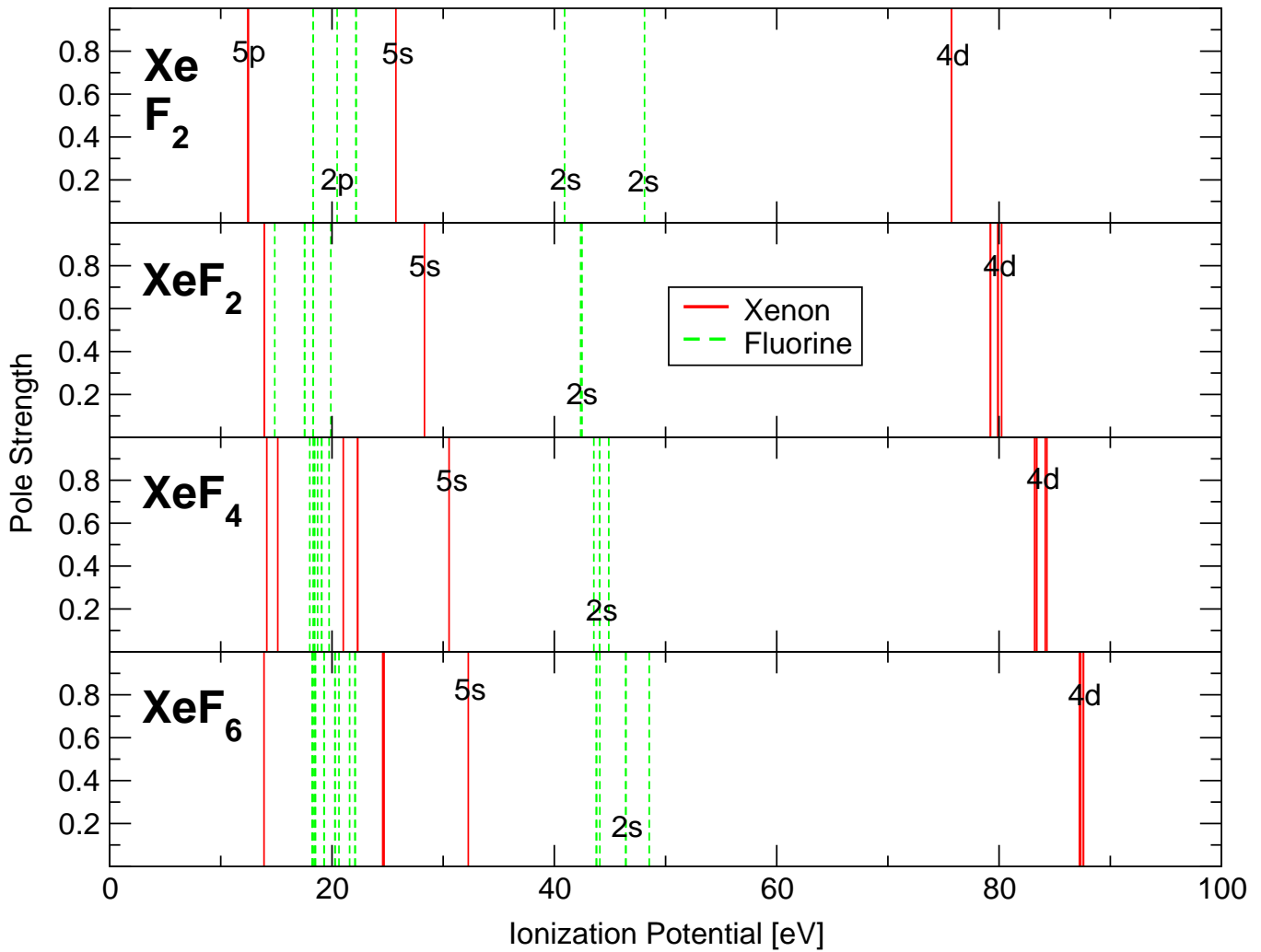}
    \caption{(Color) Single ionization spectra of~Xe, \Ft, \XeFt,
	     \XeFf, and \XeFs\ calculated at the Hartree-Fock level
	     (Koopmans' theorem). The assignment of the lines to atomic
	     orbitals of xenon or fluorine origin in~\XeFn\ is not
	     strict in the valence region, due to the molecular bond.}
    \label{fig:XeFx_HF}
  \end{center}
\end{figure*}

The mentioned shifts of the orbital energies of the xenon atom are
used to correct the Xe$\,4d$~orbital energies for the scalar relativistic
effect by adding~$-\Delta\varepsilon$ to the Xe$\, 4d$~ionization potential
of the nonrelativistic calculations. This procedure is justified by the
observation that the molecular Xe$\,4d$~orbitals are highly localized and
are very similar to atomic Xe$\,4d$~orbitals. The Xe$\,4d$~ionization
potentials, calculated with the ADC(3) scheme, take electron correlation into
account. Therefore, contributions of many orbitals mix into the
description of the Xe$\,4d$~ionized states. Since the
Xe$\,4d$~ionized states are predominantly described by the
Xe$\,4d$~orbitals, the relativistic correction is applicable in
this case as well.

The spin-orbit splitting in the xenon atom amounts to~$2.111 \U{eV}$
for the $4d$~orbitals and $1.436 \U{eV}$ for $5p$~orbitals
(see Table~\ref{tab:Xe-orbitals}). The spin-orbit splitting in the xenon
atom is probably only a good approximation~\cite{Cutler:Xe-91} for
the Xe$\, 4d$~orbitals in the xenon fluorides, because the Xe$\,
5p$~orbitals suffer from a significant modification by the
molecular bond to the fluorine atoms. The experimental value for
the spin-orbit splitting of the $4d$~orbitals in the xenon atom
is~$1.984 \pm 0.014 \U{eV}$~\cite{Cutler:Xe-91} (see Table~\ref{tab:XeFx}).
The theoretically and experimentally determined values are in
satisfactory agreement.	 A good agreement is not expected because
the Dirac-Fock equations are based on a mean-field
approximation.~\cite{Balasubramanian:RE-97} Spin-orbit splitting
is not accounted for in the nonrelativistic theory used throughout
and is not considered any further in the ensuing sections.

\section{Results and Discussion}
\label{sec:results}
\subsection{Mean-Field Model}
\label{sec:mean}

\begin{figure*}
  \begin{center}
    \includegraphics[width=\hsize,clip]{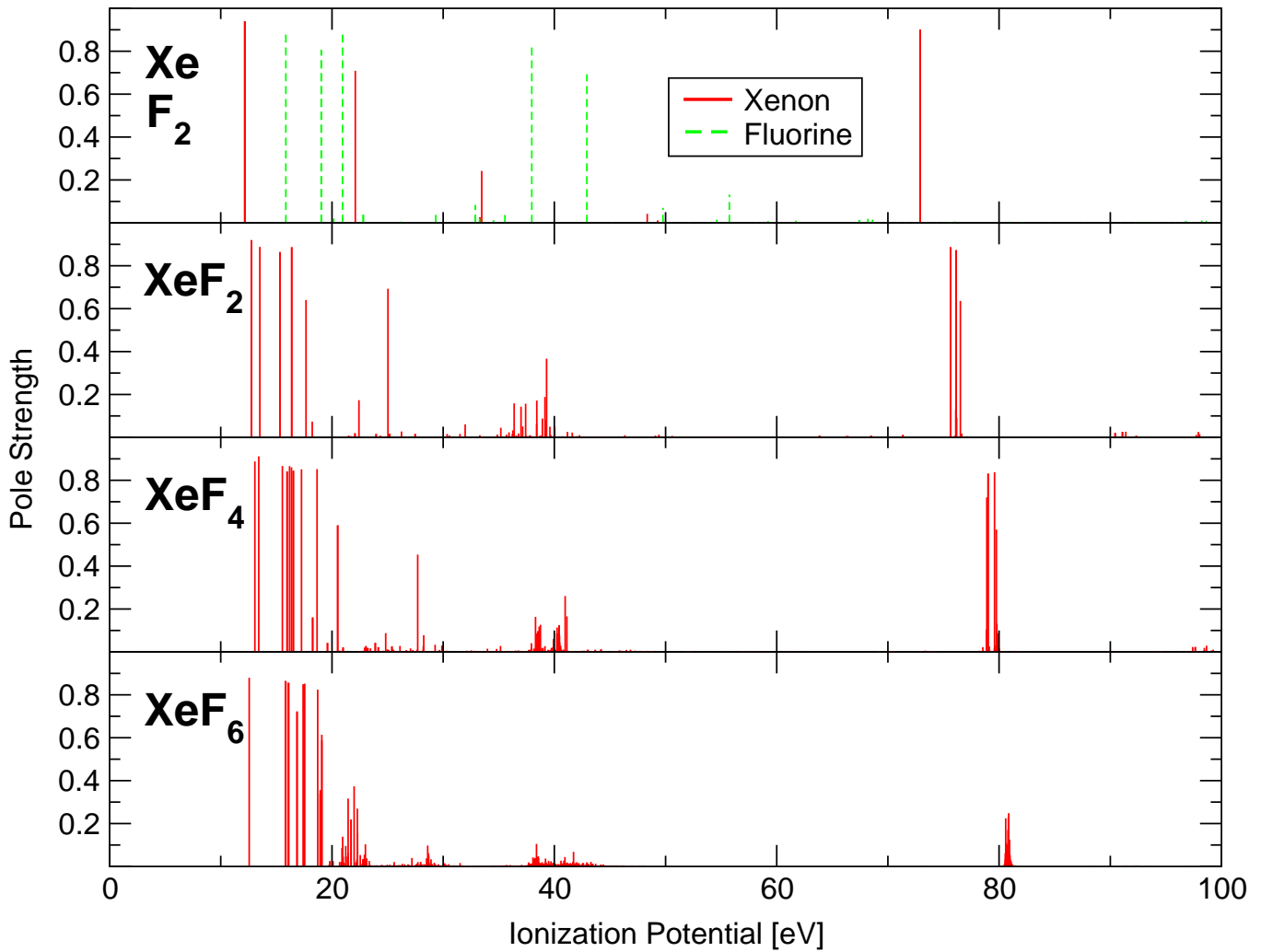}
    \caption{(Color) Single ionization spectra of~Xe, \Ft, \XeFt, \XeFf,
	     and \XeFs\ calculated with the one-particle ADC(3) program (see
	     the text). The assignment of lines to atomic orbitals of xenon or
	     fluorine origin is not done for the \XeFn~molecules.}
    \label{fig:XeFx}
  \end{center}
\end{figure*}

Single ionization spectra of the xenon fluorides can be obtained
in the Hartree-Fock approximation with the help of Koopmans'
theorem.~\cite{Szabo:MQC-82,Koopmans:-33} In this mean-field model,
the correlation between the electrons is neglected, resulting in
very simple spectra (see Fig.~\ref{fig:XeFx_HF}) that can provide hints
for the interpretation of the more complex spectra that include
electron correlation (see Fig.~\ref{fig:XeFx}). The assignment of
the lines in Fig.~\ref{fig:XeFx_HF} to atomic orbitals of xenon or
fluorine is not strict in the valence region, due to the molecular bond.
In the deeper lying molecular orbitals, this assignment is well defined.
Since spin-orbit coupling is neglected, artificial degeneracies are
introduced in the spectra of Fig.~\ref{fig:XeFx_HF}. The effect of the
ligand field caused by the fluorine atoms is of course included.

The effect of adding fluorine atoms to xenon is studied in
Table~\ref{tab:mulliken} by Mulliken and L\"owdin population
analyses.~\cite{Szabo:MQC-82,Mulliken:-55} Such population analyses
may exhibit some basis set dependence,~\cite{Szabo:MQC-82} but the
consistency between the two sets of results indicates reliability.
One sees immediately that charge is withdrawn from the xenon atom
to the fluorine atoms: $1.1$~electron charges in \XeFt, $1.9$ in \XeFf,
and $2.3$ in \XeFs. Due to the $C_{3v}$~symmetry of the
ground-state geometry of~\XeFs, there are two kinds of fluorine
atoms, with different distances to the central xenon atom. These
differences in nuclear separation are reflected in
Table~\ref{tab:mulliken} by the fact that fluorine atoms which are
further away from the xenon atom acquire less charge.

\begin{table}
  \centering
  \begin{tabular}{c|r|c|c|c}
    Compound & Atoms & Nuclear & Mulliken   & L\"owdin	 \\
	     &	     & Charge  & Population & Population \\
    \hline\hline
    \XeFt    & Xe    &		 54 &		52.92 &		  52.95 \\
	       & 2 \ F & \phantom{0}9 & \phantom{0}9.54 & \phantom{0}9.53 \\
    \hline
    \XeFf    & Xe    &		 54 &		52.13 &		  52.12 \\
	       & 4 \ F & \phantom{0}9 & \phantom{0}9.47 & \phantom{0}9.47 \\
    \hline
    \XeFs    & Xe    &		 54 &		51.71 &		  51.51 \\
	       & 3 \ F & \phantom{0}9 & \phantom{0}9.49 & \phantom{0}9.51 \\
	       & 3 \ F & \phantom{0}9 & \phantom{0}9.27 & \phantom{0}9.33 \\
  \end{tabular}
  \caption{Mulliken and L\"owdin population analyses of~\XeFt, \XeFf, and
	   \XeFs.}
  \label{tab:mulliken}
\end{table}

At first sight one might assume that the computed change in charge
density involves only the valence electrons and has little effect
on the inner molecular orbitals of the xenon fluorides.	 In the
hydrogen atom, the wave functions of the higher lying shells,
which are unoccupied in the ground state, have a considerable
amplitude in the spatial regions of the lower lying shells of the
same angular momentum.~\cite{Sakurai:MQM-94} As this argument also
holds in the case of the xenon fluorides, a reduction of valence
electron density on the xenon atom leads to a less efficient
screening of its nuclear charge and consequently to a lowering in
energy of the energetically low lying molecular orbitals with a
dominant contribution on the xenon atom.

Conversely, the increase of valence electron density on the
fluorine atoms leads to a more efficient screening of their
nuclear charge and consequently raises the energy of the lower
lying molecular orbitals with a dominant contribution on fluorine
atoms. The charge withdrawn from the xenon atom is shared among
several fluorine atoms. The net increase of charge density on each
fluorine atom is $\approx 0.5$~electron charges. This value is, of
course, much smaller than the loss of charge density on the xenon
atom and hence the impact on the inner molecular orbitals of
fluorine character is clearly much smaller than for xenon.
It is most dramatic in~\XeFt\ where each fluorine atom acquires
the largest fraction of charge. The screening of the nuclear
charge is largest in~\XeFt\ and decreases in~\XeFf\ and~\XeFs.

The effects of this model can be seen in Fig.~\ref{fig:XeFx_HF}.
The positions of the Xe$\,5s$ and Xe$\,4d$ lines shift to higher
binding energies with an increasing number of fluorine atoms.  The
energy differences between the corresponding lines of~\XeFt\ and
\XeFf\ and of~\XeFf\ and \XeFs\ are nearly equally large. The
F$\,2p$~and F$\,2s$~lines also shift slightly to higher binding
energies with increasing number of fluorine atoms. The charge
redistribution over several fluorine ligands we mentioned earlier
furnishes an explanation for the shift of the fluorine lines. This
interpretation is further supported by a comparison with~\Ft.  The
mean of the F$\,2p$~lines in~\Ft\ and the mean of the F$\,2s$~lines
in~\Ft\ are higher in energy than the mean values of the corresponding
lines in~\XeFt. The first ionization potential is nearly constant for
all xenon fluorides studied. Its value is~$\approx 12.5 \U{eV}$.

In the fluorine molecule, the two F$\,2s$~lines are split
considerably due to the molecular bond.	 The splitting of the two
F$\,2s$~lines in~\XeFt\ is tiny due to the large separation of the
fluorine atoms. In~\XeFf\ and \XeFs, the fluorine atoms are closer
and interact. This results in a larger splitting of the F$\,2s$~lines
in comparison to the splitting in~\XeFt.

The Xe$\,4d$~lines are quintuply degenerate in the single
ionization spectrum of the xenon atom in Fig.~\ref{fig:XeFx_HF}.
In~\XeFt, the degeneracy is lifted by the ligand field of the
fluorine atoms and three distinct lines become visible. The three
lines reflect the spatial orientation of the $4d$~orbitals. There
is a ligand field along the molecular axis. Perpendicular to the
molecular axis there is no shift, resulting in a total of three
lines. In~\XeFf\ there is only one dimension left that is
unaffected by the ligand field: the axis perpendicular to the
molecular plane. The spectrum, Fig.~\ref{fig:XeFx_HF}, shows
that there are two distinct groups of Xe$\,4d$~lines in this case.

In~\XeFs, the situation changes because the fluorine ligands are
grouped around the xenon atom in such a way that~\XeFs\ is close to
octahedral symmetry. $O_h$~symmetry in \XeFs\ leads to two distinct
Xe$\,4d$~orbital energies separated by only~$0.1568 \U{eV}$. Lowering
the symmetry of \XeFs\ to~$C_{3v}$ results in three distinct Xe$\,4d$~orbital
energies with a total splitting of~$0.3682 \U{eV}$, which is still small
compared to the much larger Xe$\,4d$~splittings in~\XeFt\ and \XeFf.

\subsection{Correlation Effects}

Going beyond the Hartree-Fock description of the molecules by
using~ADC(3) leads to the ionization potentials plotted in
Fig.~\ref{fig:XeFx}. Figure~\ref{fig:XeFx_HF} helps identify the
one-particle origin of the states in Fig.~\ref{fig:XeFx}.
The Xe$\,4d$~lines are energetically clearly separated from the
outer and inner valence lines in all spectra.  They are located
between~$72 \U{eV}$ and $82 \U{eV}$. The F$\,2s$-derived~lines are present
between $35 \U{eV}$ and $45 \U{eV}$. Correlation effects do not change
the fact that the lowest ionization potentials are approximately equal
in all compounds.

\renewcommand{\arraystretch}{1.2}%
\begin{table}
  \centering
  \begin{tabular}{c|c|c}
    Compound & First Experimental IP & First ADC IP \\
    \hline\hline
    Xe	   & 12.129 & 12.16 \\
    \hline
    \XeFt  & 12.35  & 12.76 \\
    \hline
    \XeFf  & 13.1   & 13.07 \\
    \hline
    \XeFs  & 12.35  & 12.56 \\
  \end{tabular}
  \caption{Comparison of the calculated lowest (first) ionization
	   potentials~(IP) of~Xe, \XeFt, \XeFf, and \XeFs\ with experimental
	   results. The first ionization potential of xenon is taken
	   from Ref.~\onlinecite{Holleman:IC-01}. The other first ionization
	   potentials are taken from Ref.~\onlinecite{Nielsen:XeF-76}.
	   All data are given in electronvolt.}
  \label{tab:IPvalence}
\end{table}
\renewcommand{\arraystretch}{1}%

In fact, at the lowest ionization potentials, which are associated with the
Xe$\,5p$ levels, there is qualitative agreement between the correlated and
the mean-field calculation: The pole strengths are very close to unity;
the one-particle picture is valid. Nevertheless, ADC(3) improves on the
Hartree-Fock method in a quantitative manner. In the correlated calculation,
the energetically lowest cations are stabilized by about $1 \U{eV}$ relative
to the respective one-particle energies. The first ionization potentials of
xenon and its fluorides, as calculated within the ADC(3)~scheme, are compared
to the experimental results in Table~\ref{tab:IPvalence}. The agreement of
experimental and calculated ionization potentials is good.

At higher energies, there is typically not even qualitative agreement between
mean-field and many-body treatment. Overall, there is a dramatic
change of the ionization spectra due to the inclusion of many-body effects.
The importance of electron correlation grows with increasing number of
fluorine atoms. This is reflected in energy shifts and reduction of intensity
in the outer valence part of the spectra; in the appearance of numerous
satellite lines in the vicinity of~$20 \U{eV}$ and above; and, in particular,
in the eye-catching breakdown of the molecular orbital picture of ionization
seen in the Xe$\,5s$ and F$\,2s$~inner-valence region.

The latter phenomenon is common in molecules:~\cite{Cederbaum:CE-86} The
intensity originally confined to a single orbital in Fig.~\ref{fig:XeFx_HF}
is spread over many cationic states (see Fig.~\ref{fig:XeFx}). The breakdown
of the molecular orbital picture of ionization is caused by compact
two-hole--one-particle configurations that are close in energy to a
one-hole state.~\cite{Cederbaum:CE-86} The strong Coulomb coupling
of these configurations to one-hole states leads to a broad shape of
lines, which has been discovered in the inner-valence region of many
molecules.~\cite{Cederbaum:CE-86}

Due to the many-body description of a molecule by the ADC~scheme,
decay processes of electronic states of the ion that lie above the
double ionization threshold also become describable.~\cite{Zobeley:HE-98} In
particular, the final state of the electronic decay of an ionized
molecule can be approximated in terms of two-hole--one-particle
configurations describing the emitted electron~(the particle) and the residual
dication~(the two holes).~\cite{Santra:ED-01} The shape of each decaying
state can be identified as a thin bundle of lines, which mimic a discretized
Lorentzian curve.~\cite{Zobeley:HE-98}

In order to unambiguously identify a decaying state, the breakdown of the
molecular orbital picture of ionization has to be separated
clearly from the decaying states in a small energy range where
each decaying state is represented by a discretized Lorentzian curve.
Furthermore, there can be a mixing of both phenomena for a certain
state, but it is hard to identify a decay curve for one-hole
states that suffer from the breakdown phenomenon.

The inner-valence region of the xenon fluorides comprises
the~Xe$\,5s$ and F$\,2s$~states. The F$\,2s$~states are subject to
the breakdown phenomenon and can also decay by emitting an
electron.~\cite{Buth:NH-02,Buth:-03} The Xe$\,5s$~one-hole state
suffers solely from breakdown. No decay is possible because its ionization
potential is below the double ionization threshold.~\cite{Buth:NH-02,Buth:-03}
Many two-hole--one-particle configurations can couple to the
Xe$\,5s$~one-hole state (see Fig.~\ref{fig:XeFx}). As the
Xe$\,5s$~line is close to the double ionization
threshold,~\cite{Buth:NH-02,Buth:-03} many two-hole--one-particle
configurations are close in energy to this state: The excited states
are generally very dense in this energetic region.

The spectra exhibit a considerable increase of breakdown of the
molecular orbital picture for the inner valence with an increasing
number of fluorine atoms. Obviously, the number of suitable
configurations rises due to the addition of fluorine atoms. Since
xenon fluorides are very symmetric, these fluorine atoms are all
equivalent (except for~\XeFs\ in $C_{3v}$~symmetry), and the number
of equivalent two-hole--one-particle configurations with one hole on
the xenon atom and one hole on a fluorine atom doubles between~\XeFt\
and~\XeFf. In~\XeFs\ there are even more such configurations than
there are in~\XeFf, but they are no longer equivalent due to the
$C_{3v}$~symmetry of the \XeFs~ground state.

Another contribution to the increase of one-hole states
involved in the breakdown phenomenon can be attributed to the
decrease in symmetry: spherical symmetry~(Xe), $D_{\infty
h}$~(\XeFt), $D_{4h}$~(\XeFf), $C_{3v}$~(\XeFs). This decrease in
symmetry leads to an increase in excited electronic configurations that
can couple to the respective one-hole states.

The one-particle picture provides a rather good description for the
Xe$\,4d$~lines. Two-hole--one-particle configurations close to the energy
of the core one-hole configurations are spatially diffuse and thus couple
only weakly to the core one-hole configurations. Therefore, in the core regime,
the mixing of two-hole--one-particle configurations with one-hole
configurations is relatively weak. The splittings of the Xe$\,4d$~lines in
the xenon fluorides are of comparable size to those in Fig.~\ref{fig:XeFx_HF},
and the individual lines can be identified easily in all molecules but~\XeFs.

Since in all xenon fluorides the Xe$\,4d$~lines are above the double
ionization thresholdi,~\cite{Buth:NH-02,Buth:-03} Xe$\,4d$-ionized xenon
fluorides can undergo Auger decay. The Xe$\,4d$~lines in Fig.~\ref{fig:XeFx} 
show signs of decay by exhibiting features of discretized Lorentzian 
curves. The shape of the approximate Lorentzian curve representing a 
decaying state in the spectra depends highly on the number of 
two-hole--one-particle configurations in the energy range of the 
decay electron.

The decay electrons that result from the decay of a Xe$\,4d$~vacancy
are highly energetic (up to~$\approx 60 \U{eV}$). Hence, a satisfactory
description in terms of Gaussian basis sets is not feasible. For this reason,
the Xe$\,4d$~ionization potentials in Fig.~\ref{fig:XeFx} correspond in
our description essentially to one-hole states (a pole strength close to
unity). Nevertheless, the number of states which originate from
two-hole--one-particle configurations around the Xe$\,4d$ ionization
potentials grows substantially with the number of fluorine atoms.
In~\XeFs\ one observes very densely lying states around~$83 \U{eV}$
with a high contribution of two-hole--one-particle configurations. The
extreme change in the importance of two-hole--one-particle configurations
in~\XeFs\ compared to~\XeFf, \XeFt\ was investigated by us.

The six fluorine atoms in~\XeFs\ form a distorted octahedron
around the xenon atom. The basis set of the fluorine atoms may therefore
help improve the description of the decay electron from
Xe$\,4d$-ionized \XeFs\ considerably. To test this assumption, we
calculated the ionization spectrum of~\XeFt\ with the DZVP (DFT orbital)
basis set augmented by a few diffuse functions. The resulting spectrum did not
change much compared to that of Fig.~\ref{fig:XeFx}.

To test the effect of the basis functions on the fluorine atoms,
we performed another calculation of the ionization spectrum
of~\XeFt\ with fluorine basis functions attached to ghost centers
arranged such as to yield an octahedron of fluorine basis functions
surrounding the central xenon atom. The extra basis functions had
a minor effect on the spectrum and did not account for the drastic
effect observed in Fig.~\ref{fig:XeFx} for~\XeFs.

These investigations convinced us of the suitability of the DZVP
(DFT orbital) basis set for our concerns but did not clarify the reasons
for the drastic change in Fig.~\ref{fig:XeFx}. To this end, we calculated
the ionization spectrum of~\XeFs\ in $O_h$~geometry. The resulting
Xe$\,4d$~lines had a shape similar to that of the Xe$\,4d$~lines
in~\XeFf, revealing the great importance of the
decreased symmetry for the description of~\XeFs\ in our computations. The
reduction in symmetry from~$O_h$ to~$C_{3v}$ enables a vastly enlarged
amount of two-hole--one-particle configurations to couple to the Xe$\,4d$
one-hole configurations. To some extent, this is similar to an increased
basis set, as a multitude of configurations are provided additionally
for the description of the ionized states. For sure, this improves the
description of those decay electrons with low kinetic energy.

The charge transfer from the xenon atom to the fluorine atoms in the
ground state, discussed in Sec.~\ref{sec:mean}, leaves a positively
charged central xenon atom. Therefore, it is not surprising
that~\XeFf\ and \XeFs\ possess positive electron affinities,
\ie, (\XeFf)$^{-}$ and (\XeFs)$^{-}$ are electronically stable anions.
We have computed the respective electron affinities by ADC(3). The value
obtained for \XeFf\ (a doubly degenerate state) is~$0.66 \U{eV}$;
the electron affinities of~\XeFs\ lie at~$1.30 \U{eV}$, $1.31 \U{eV}$, and
$1.81 \U{eV}$, respectively. The figures are presumably not very accurate
because the DZVP (DFT orbital) basis set is not particularly well suited for
describing the diffuse states of anions. In~\XeFs, the fluorine atoms form a
cage of negative charge, of nearly octahedral symmetry, surrounding the
central, positively charged xenon atom. The resulting potential well is
similar to the model potential well discussed in
Refs.~\onlinecite{Santra:NH-02,Buth:NH-02}. In these references, both bound
states as well as scattering resonances in the model potential well were
investigated.

\renewcommand{\arraystretch}{1.2}%
\begin{table}
  \centering
  \begin{tabular}{l|c|l|l|l|l}
    Cmpd. & Line & \hfil IP$\I{expt}$ & \hfil IP$\I{ADC}$
		 & \hfil IP$\I{rel }$ & \hfil $\Gamma\I{expt}$ \\
    \hline\hline
    Xe	  & 1 & 69.525 (10) & 72.90 & 69.82 & 0.207 (4)	 \\
	  & 2 & 67.541 (9)  &	    &	    & 0.202 (4)	 \\
    \hline
    \XeFt & 1 & 72.568 (6)  & 76.52 & 73.44 & 0.248 (8)	 \\
	  & 2 & 72.248 (6)  & 76.14 & 73.06 & 0.223 (10) \\
	  & 3 & 70.601 (13) & 75.63 & 72.55 & 0.264 (26) \\
	  & 4 & 70.421 (9)  &	    &	    & 0.256 (27) \\
	  & 5 & 70.179 (6)  &	    &	    & 0.214 (19) \\
    \hline
    \XeFf & 1 & 75.098 (6)  & 79.76 & 76.68 & 0.319 (8)	 \\
	  & 2 & 74.729 (7)  & 79.60 & 76.52 & 0.255 (8)	 \\
	  & 3 & 73.140 (10) & 79.01 & 75.93 & 0.392 (10) \\
	  & 4 & 72.816 (10) & 78.89 & 75.81 & 0.210 (27) \\
	  & 5 & 72.661 (5)  &	    &	    & 0.225 (26) \\
    \hline
    \XeFs & 1 & 77.462 (13) & 80.86 & 77.78 & 0.32 (4)	 \\
	  & 2 & 77.321 (11) & 80.59 & 77.51 & 0.25 (3)	 \\
	  & 3 & 75.53	    &	    &	    & 0.33	 \\
	  & 4 & 75.38	    &	    &	    & 0.25	 \\
	  & 5 & 75.25	    &	    &	    & 0.25	 \\
  \end{tabular}
  \caption{Peak positions and widths of the~Xe$\,4d$~lines in~Xe,
	   \XeFt, \XeFf, and \XeFs. \textit{Labels}---Cmpd: compound; 
           IP$\I{expt}$: experimental peak position; IP$\I{ADC}$: 
           calculated peak position; IP$\I{rel}$:~calculated peak 
           position with relativistic correction; $\Gamma\I{expt}$: 
           experimental peak width.  IP$\I{expt}$~and $\Gamma\I{expt}$ are
	   photoelectron-experimental data, reproduced from Table~1~in
	   Ref.~\onlinecite{Cutler:Xe-91}. For~\XeFs, the data with an
	   experimental resolution of~$0.11 \U{eV}$ are taken.  The values in
	   parentheses are the standard deviations of the peak positions and
	   peak widths.~\cite{Cutler:Xe-91} IP$\I{ADC}$ and
	   IP$\I{rel}$~are sorted descending in energy for each compound.
	   The value of~IP$\I{expt}$ does not necessarily correspond to
	   the values of~IP$\I{ADC}$ and IP$\I{rel}$ in the
	   same row as spin-orbit splitting is neglected to obtain the latter
	   ones. All data are given in electronvolt.}
  \label{tab:XeFx}
\end{table}
\renewcommand{\arraystretch}{1}%

The Xe$\,4d$ and F$\,2s$~lines in Fig.~\ref{fig:XeFx} are shifted to
higher ionization potentials with increasing number of fluorine atoms
(see Sec.~\ref{sec:mean}). This general trend, seen in the mean-field
approximation of Fig.~\ref{fig:XeFx_HF}, can still be
identified if electron correlation is taken into account.
Nevertheless, the difference between the shifts of the
Xe$\,4d$~lines of~\XeFt\ and \XeFf\ is larger than that between
the Xe$\,4d$~lines of~\XeFf\ and \XeFs. Electron correlation
reduces the effect caused by charge depletion on the xenon atom.

The nonrelativistic IP$\I{ADC}$ are listed in
Table~\ref{tab:XeFx} together with the IP$\I{rel}$ obtained
by applying to IP$\I{ADC}$ the relativistic
correction~$-\Delta\varepsilon$ of Sec.~\ref{sec:relativistic}. The number
of distinct Xe$\,4d$~lines is smaller in the nonrelativistic
spectra due to a higher degeneracy caused by neglecting the
spin-orbit coupling. The Xe$\, 4d$~lines of~\XeFs\ cannot be
identified clearly in Fig.~\ref{fig:XeFx}. One expects from
Sec.~\ref{sec:mean} two lines for a ground-state geometry of
$O_h$~symmetry and three lines for a ground-state geometry of
$C_{3v}$~symmetry. The two ionization potentials, in the energy range of the
Xe$\,4d$~lines of \XeFs~(80--83$\U{eV}$), with maximum pole
strength are listed in Table~\ref{tab:XeFx}.

The Xe$\,4d$ IP$\I{rel}$ differ from the experimentally
obtained data in Table~\ref{tab:XeFx} by only~$1.5$--$2.5 \U{eV}$
($\approx 3\%$), which is a good agreement in view of the
complexity of the problem and the necessity to describe the decay
simultaneously. The reason for the deviation is twofold. First,
the spin-orbit splitting is neglected, which amounts to~$2.111
\U{eV}$ for the Xe$\,4d$~lines of the xenon atom, according to
Sec.~\ref{sec:relativistic}. Second, since the calculations are based on
the ADC(3) approximation, the three-hole--two-particle configurations
are neglected [ADC(4) calculations are currently beyond reach]. The
inclusion of these additional configurations would shift the
ionization potentials of the Xe$\, 4d$~lines further to lower energy,
due to an improved description of core-hole relaxation.~\cite{Angonoa:KS-87}

Let us now turn our attention to the last column in Table~\ref{tab:XeFx}:
the experimental widths of the Xe$\, 4d$~photoelectron
peaks.~\cite{Cutler:Xe-91} It is evident that, on average, the widths
increase from Xe to XeF$_6$. According to Cutler~\etal,~\cite{Cutler:Xe-91} 
this effect probably cannot be explained in terms of vibrational 
broadening. If the data reflect more or less pure
Auger widths, then the experimental findings are in contradiction to
expectations based on the one-center model.~\cite{Coville:ME-91} The
electronegative fluorine ligands withdraw charge from xenon (see
Table~\ref{tab:mulliken}). Correspondingly, fewer valence electrons
are available for a local, atomic mechanism of Xe$\, 4d$~Auger decay.
Hartmann's semiempirical multi-center model,~\cite{Hartmann:MC-88} on the
other hand, predicts a larger C$\, 1s$~Auger width for CF$_4$ than for
CH$_4$, which, though seemingly in line with the trend in \XeFn,
is inconsistent with experiment.~\cite{Carroll:PE-02}

In our work on neon clusters, we demonstrated not only that a Ne$\, 2s$~hole
\emph{in a cluster} can decay by electron emission~\cite{Santra:ICD-00}---a fact
confirmed by a recent experiment~\cite{Marburger:EE-03}---but also that
the decay lifetime depends sensitively on the number of nearest neighbors
surrounding the atom carrying the inner-valence hole.~\cite{Santra:ED-01} The
Ne$\, 2s$~lifetime drops in a monotonic manner from about $80 \U{fs}$ in
Ne$_2^+$ to less than $5 \U{fs}$ in Ne$_{13}^+$. The mechanism responsible
for this dramatic effect is referred to as \emph{interatomic Coulombic decay}.
In heteroclusters, additional electronic mechanisms play a
role.~\cite{Zobeley:ED-01} We will demonstrate in an upcoming
paper~\cite{Buth:-03} how the decay mechanisms discovered in clusters
elucidate the counterintuitive behavior of the Xe$\, 4d$~Auger widths
in \XeFn.

Here, we would like to emphasize that traces of a purely electronic origin
can already be found in our ADC(3) spectra, Fig.~\ref{fig:XeFx}. As the
number of fluorine ligands grows, there are more and more
two-hole--one-particle configurations that can couple to the Xe$\, 4d$~holes.
In other words, the number of dicationic decay channels grows, and therefore
we observe in Fig.~\ref{fig:XeFx} an increasing loss of main-line intensity
and the emergence of dense line bundles. This is the signature in our
finite-basis-set calculations of enhanced electronic
decay,~\cite{Zobeley:HE-98} as pointed out previously.

\section{Conclusion}
\label{sec:conclusion}

The xenon fluorides are very special systems. Not only are they rather
peculiar chemical compounds, they are also particularly well suited to
investigating the systematic evolution of many-body effects.

While outer-valence ionization is mostly compatible with the mean-field
approach, the situation is strikingly different in the inner-valence
region. At the Hartree-Fock level, the chemical environment leads to mere
shifts of the Xe$\, 5s$ and F$\, 2s$~lines. The ADC(3) calculation, however,
reveals that electron correlation can induce additional, pronounced
environmental effects. Xe$\, 5s$ ionization in XeF$_2$ can be relatively well
understood in terms of a one-particle picture. In XeF$_4$, this picture begins
to crumble, and in XeF$_6$, there is not a single cationic eigenstate that
overlaps well with the Xe$\, 5s$~one-hole configuration.

The F$\, 2s$~states are even more interesting, since they are higher in energy
than the double ionization threshold. In the xenon fluorides, electron
correlation enforces the excitation of a large, system-dependent number
of F$\, 2s$-derived electronic states. All of these states are resonances
decaying by electron emission. Neither the decay mechanism nor the decay
time scale are known. Resolving these issues may have important consequences
for other molecules. This poses an exciting challenge to both theory and
experiment.

This is the first paper dealing with the inner-valence physics of \XeFn\
at a correlated level. We are not aware of any experimental spectra in this
energy region. The quality of our calculations can be assessed, however, by
reference to our ADC(3) results for the Xe$\, 4d$~core lines. Upon correction
for scalar relativistic effects, our numerical data agree with experiment
within a few percent. The remaining difference can be attributed to an
incomplete description of core-hole relaxation and to our neglect of
spin-orbit coupling. The ADC(3) scheme is capable of reproducing the
experimentally observed line shifts as well as the counterintuitive Auger
broadening. A detailed analysis of the latter phenomenon will be presented
elsewhere.

\begin{acknowledgments}
We are highly indebted to T.~Darrah Thomas for drawing our attention to
the xenon fluorides. He pointed out Ref.~\onlinecite{Cutler:Xe-91} and
supported it further with valuable private communications. Markus
Pernpointner helped estimate the impact of the relativistic effects
in the xenon atom. This work would not have been possible without the
ADC~programs and support by Francesco Tarantelli. Imke B. M\"uller,
Sven Feuerbacher, J\"org Breidbach, and Thomas Sommerfeld
accompanied our work with helpful comments and fruitful discussions.
R.~S.~and L.~S.~C. gratefully acknowledge financial
support by the Deutsche Forschungsgemeinschaft~(DFG).
\end{acknowledgments}


\begin{thebibliography}{49}
\expandafter\ifx\csname natexlab\endcsname\relax\def\natexlab#1{#1}\fi
\expandafter\ifx\csname bibnamefont\endcsname\relax
  \def\bibnamefont#1{#1}\fi
\expandafter\ifx\csname bibfnamefont\endcsname\relax
  \def\bibfnamefont#1{#1}\fi
\expandafter\ifx\csname citenamefont\endcsname\relax
  \def\citenamefont#1{#1}\fi
\expandafter\ifx\csname url\endcsname\relax
  \def\url#1{\texttt{#1}}\fi
\expandafter\ifx\csname urlprefix\endcsname\relax\def\urlprefix{URL }\fi
\providecommand{\bibinfo}[2]{#2}
\providecommand{\eprint}[2][]{\url{#2}}

\bibitem[{\citenamefont{Wiberg et~al.}(2001)\citenamefont{Wiberg, Holleman, and
  Wiberg}}]{Holleman:IC-01}
\bibinfo{author}{\bibfnamefont{N.}~\bibnamefont{Wiberg}},
  \bibinfo{author}{\bibfnamefont{A.~F.} \bibnamefont{Holleman}},
  \bibnamefont{and} \bibinfo{author}{\bibfnamefont{E.}~\bibnamefont{Wiberg}},
  \emph{\bibinfo{title}{Inorganic Chemistry}} (\bibinfo{publisher}{Academic
  Press}, \bibinfo{address}{New York}, \bibinfo{year}{2001}), ISBN
  \bibinfo{isbn}{0-123-52651-5}.

\bibitem[{\citenamefont{Comes et~al.}(1973)\citenamefont{Comes, Haensel,
  Nielsen, and Schwarz}}]{Comes:XeF-73}
\bibinfo{author}{\bibfnamefont{F.~J.} \bibnamefont{Comes}},
  \bibinfo{author}{\bibfnamefont{R.}~\bibnamefont{Haensel}},
  \bibinfo{author}{\bibfnamefont{U.}~\bibnamefont{Nielsen}}, \bibnamefont{and}
  \bibinfo{author}{\bibfnamefont{W.~H.~E.} \bibnamefont{Schwarz}},
  \bibinfo{journal}{J. Chem. Phys.} \textbf{\bibinfo{volume}{58}},
  \bibinfo{pages}{516} (\bibinfo{year}{1973}).

\bibitem[{\citenamefont{Nielsen et~al.}(1974)\citenamefont{Nielsen, Haensel,
  and Schwarz}}]{Nielsen:XeF-74}
\bibinfo{author}{\bibfnamefont{U.}~\bibnamefont{Nielsen}},
  \bibinfo{author}{\bibfnamefont{R.}~\bibnamefont{Haensel}}, \bibnamefont{and}
  \bibinfo{author}{\bibfnamefont{W.~H.~E.} \bibnamefont{Schwarz}},
  \bibinfo{journal}{J. Chem. Phys.} \textbf{\bibinfo{volume}{61}},
  \bibinfo{pages}{3581} (\bibinfo{year}{1974}).

\bibitem[{\citenamefont{Nielsen and Schwarz}(1976)}]{Nielsen:XeF-76}
\bibinfo{author}{\bibfnamefont{U.}~\bibnamefont{Nielsen}} \bibnamefont{and}
  \bibinfo{author}{\bibfnamefont{W.~H.~E.} \bibnamefont{Schwarz}},
  \bibinfo{journal}{Chem. Phys.} \textbf{\bibinfo{volume}{13}},
  \bibinfo{pages}{195} (\bibinfo{year}{1976}).

\bibitem[{\citenamefont{Carroll et~al.}(1974)\citenamefont{Carroll, Shaw~Jr.,
  Thomas, Kindle, and Bartlett}}]{Carroll:XeF-74}
\bibinfo{author}{\bibfnamefont{T.~X.} \bibnamefont{Carroll}},
  \bibinfo{author}{\bibfnamefont{R.~W.} \bibnamefont{Shaw~Jr.}},
  \bibinfo{author}{\bibfnamefont{T.~D.} \bibnamefont{Thomas}},
  \bibinfo{author}{\bibfnamefont{C.}~\bibnamefont{Kindle}}, \bibnamefont{and}
  \bibinfo{author}{\bibfnamefont{N.}~\bibnamefont{Bartlett}},
  \bibinfo{journal}{J. Am. Chem. Soc.} \textbf{\bibinfo{volume}{96}},
  \bibinfo{pages}{1989} (\bibinfo{year}{1974}).

\bibitem[{\citenamefont{Cutler et~al.}(1991)\citenamefont{Cutler, Bancroft,
  Bozek, Tan, and Schrobilgen}}]{Cutler:Xe-91}
\bibinfo{author}{\bibfnamefont{J.~N.} \bibnamefont{Cutler}},
  \bibinfo{author}{\bibfnamefont{G.~M.} \bibnamefont{Bancroft}},
  \bibinfo{author}{\bibfnamefont{J.~D.} \bibnamefont{Bozek}},
  \bibinfo{author}{\bibfnamefont{K.~H.} \bibnamefont{Tan}}, \bibnamefont{and}
  \bibinfo{author}{\bibfnamefont{G.~J.} \bibnamefont{Schrobilgen}},
  \bibinfo{journal}{J. Am. Chem. Soc.} \textbf{\bibinfo{volume}{113}},
  \bibinfo{pages}{9125} (\bibinfo{year}{1991}).

\bibitem[{\citenamefont{Basch et~al.}(1971)\citenamefont{Basch, Moskowitz,
  Hollister, and Hankin}}]{Basch:XeF-71}
\bibinfo{author}{\bibfnamefont{H.}~\bibnamefont{Basch}},
  \bibinfo{author}{\bibfnamefont{J.~W.} \bibnamefont{Moskowitz}},
  \bibinfo{author}{\bibfnamefont{C.}~\bibnamefont{Hollister}},
  \bibnamefont{and} \bibinfo{author}{\bibfnamefont{D.}~\bibnamefont{Hankin}},
  \bibinfo{journal}{J. Chem. Phys.} \textbf{\bibinfo{volume}{55}},
  \bibinfo{pages}{1922} (\bibinfo{year}{1971}).

\bibitem[{\citenamefont{Scheire et~al.}(1980)\citenamefont{Scheire, Phariseau,
  Nuyts, Foti, and Smith~Jr.}}]{Scheire:XeF-80}
\bibinfo{author}{\bibfnamefont{L.}~\bibnamefont{Scheire}},
  \bibinfo{author}{\bibfnamefont{P.}~\bibnamefont{Phariseau}},
  \bibinfo{author}{\bibfnamefont{R.}~\bibnamefont{Nuyts}},
  \bibinfo{author}{\bibfnamefont{A.~E.} \bibnamefont{Foti}}, \bibnamefont{and}
  \bibinfo{author}{\bibfnamefont{V.~H.} \bibnamefont{Smith~Jr.}},
  \bibinfo{journal}{Physica~A} \textbf{\bibinfo{volume}{101}},
  \bibinfo{pages}{22} (\bibinfo{year}{1980}).

\bibitem[{\citenamefont{Gutsev and Smoijar}(1981)}]{Gutsev:XeF-81}
\bibinfo{author}{\bibfnamefont{G.~L.} \bibnamefont{Gutsev}} \bibnamefont{and}
  \bibinfo{author}{\bibfnamefont{A.~E.} \bibnamefont{Smoijar}},
  \bibinfo{journal}{Chem. Phys.} \textbf{\bibinfo{volume}{56}},
  \bibinfo{pages}{189} (\bibinfo{year}{1981}).

\bibitem[{\citenamefont{Kaupp et~al.}(1996)\citenamefont{Kaupp, van W\"ullen,
  Franke, Schmitz, and Kutzelnigg}}]{Kaupp:Xe-96}
\bibinfo{author}{\bibfnamefont{M.}~\bibnamefont{Kaupp}},
  \bibinfo{author}{\bibfnamefont{C.}~\bibnamefont{van W\"ullen}},
  \bibinfo{author}{\bibfnamefont{R.}~\bibnamefont{Franke}},
  \bibinfo{author}{\bibfnamefont{F.}~\bibnamefont{Schmitz}}, \bibnamefont{and}
  \bibinfo{author}{\bibfnamefont{W.}~\bibnamefont{Kutzelnigg}},
  \bibinfo{journal}{J. Am. Chem. Soc.} \textbf{\bibinfo{volume}{118}},
  \bibinfo{pages}{11939} (\bibinfo{year}{1996}).

\bibitem[{\citenamefont{Styszy\'nski et~al.}(1997)\citenamefont{Styszy\'nski,
  Cao, Malli, and Visscher}}]{DFB:XeF-97}
\bibinfo{author}{\bibfnamefont{J.}~\bibnamefont{Styszy\'nski}},
  \bibinfo{author}{\bibfnamefont{X.}~\bibnamefont{Cao}},
  \bibinfo{author}{\bibfnamefont{G.~L.} \bibnamefont{Malli}}, \bibnamefont{and}
  \bibinfo{author}{\bibfnamefont{L.}~\bibnamefont{Visscher}},
  \bibinfo{journal}{J. Comput. Chem.} \textbf{\bibinfo{volume}{18}},
  \bibinfo{pages}{601} (\bibinfo{year}{1997}).

\bibitem[{\citenamefont{Liao and Zhang}(1998)}]{Liao:XeF-98}
\bibinfo{author}{\bibfnamefont{M.-S.} \bibnamefont{Liao}} \bibnamefont{and}
  \bibinfo{author}{\bibfnamefont{Q.-E.} \bibnamefont{Zhang}},
  \bibinfo{journal}{J. Phys. Chem.~A} \textbf{\bibinfo{volume}{102}},
  \bibinfo{pages}{10647} (\bibinfo{year}{1998}).

\bibitem[{\citenamefont{Fetter and Walecka}(1971)}]{Fetter:MP-71}
\bibinfo{author}{\bibfnamefont{A.~L.} \bibnamefont{Fetter}} \bibnamefont{and}
  \bibinfo{author}{\bibfnamefont{J.~D.} \bibnamefont{Walecka}},
  \emph{\bibinfo{title}{Quantum Theory of Many-Particle Systems}},
  International Series in Pure and Applied Physics, edited by Leonard I. Schiff
  (\bibinfo{publisher}{McGraw-Hill}, \bibinfo{address}{New York},
  \bibinfo{year}{1971}).

\bibitem[{\citenamefont{Mattuck}(1976)}]{Mattuck:FD-76}
\bibinfo{author}{\bibfnamefont{R.~D.} \bibnamefont{Mattuck}},
  \emph{\bibinfo{title}{A Guide to Feynman Diagrams in the Many-Body Problem}}
  (\bibinfo{publisher}{McGraw-Hill}, \bibinfo{address}{New York},
  \bibinfo{year}{1976}), \bibinfo{edition}{2nd} ed., ISBN
  \bibinfo{isbn}{0-07-040954-4}.

\bibitem[{\citenamefont{Szabo and Ostlund}(1982)}]{Szabo:MQC-82}
\bibinfo{author}{\bibfnamefont{A.}~\bibnamefont{Szabo}} \bibnamefont{and}
  \bibinfo{author}{\bibfnamefont{N.~S.} \bibnamefont{Ostlund}},
  \emph{\bibinfo{title}{Modern Quantum Chemistry: Introduction to Advanced
  Electronic Structure Theory}} (\bibinfo{publisher}{Macmillan},
  \bibinfo{address}{New York}, \bibinfo{year}{1982}), ISBN
  \bibinfo{isbn}{0-02-949710-8}.

\bibitem[{\citenamefont{Gross et~al.}(1991)\citenamefont{Gross, Runge, and
  Heinonen}}]{Gross:MP-91}
\bibinfo{author}{\bibfnamefont{E.~K.~U.} \bibnamefont{Gross}},
  \bibinfo{author}{\bibfnamefont{E.}~\bibnamefont{Runge}}, \bibnamefont{and}
  \bibinfo{author}{\bibfnamefont{O.}~\bibnamefont{Heinonen}},
  \emph{\bibinfo{title}{Many-Particle Theory}} (\bibinfo{publisher}{Adam
  Hilger}, \bibinfo{address}{Bristol}, \bibinfo{year}{1991}), ISBN
  \bibinfo{isbn}{0-7503-0155-4}.

\bibitem[{\citenamefont{Cederbaum and Domcke}(1977)}]{Cederbaum:TA-77}
\bibinfo{author}{\bibfnamefont{L.~S.} \bibnamefont{Cederbaum}}
  \bibnamefont{and} \bibinfo{author}{\bibfnamefont{W.}~\bibnamefont{Domcke}},
  in \emph{\bibinfo{booktitle}{Adv. Chem. Phys.}}, edited by
  \bibinfo{editor}{\bibfnamefont{I.}~\bibnamefont{Prigogine}} \bibnamefont{and}
  \bibinfo{editor}{\bibfnamefont{S.~A.} \bibnamefont{Rice}}
  (\bibinfo{publisher}{John Wiley \& Sons}, \bibinfo{address}{New York},
  \bibinfo{year}{1977}), vol.~\bibinfo{volume}{36}, pp.
  \bibinfo{pages}{205--344}.

\bibitem[{\citenamefont{Cederbaum}(1998)}]{Cederbaum:GF-98}
\bibinfo{author}{\bibfnamefont{L.~S.} \bibnamefont{Cederbaum}}, in
  \emph{\bibinfo{booktitle}{Encyclopedia of Computational Chemistry}}, edited
  by \bibinfo{editor}{\bibfnamefont{P.~v.~R.} \bibnamefont{Schleyer}}
  (\bibinfo{publisher}{John Wiley \& Sons}, \bibinfo{address}{Chichester, New
  York}, \bibinfo{year}{1998}), vol.~\bibinfo{volume}{2}, pp.
  \bibinfo{pages}{1202--1211}, ISBN \bibinfo{isbn}{0-471-96588-X}.

\bibitem[{\citenamefont{Schirmer}(1982)}]{Schirmer:PP-82}
\bibinfo{author}{\bibfnamefont{J.}~\bibnamefont{Schirmer}},
  \bibinfo{journal}{Phys. Rev.~A} \textbf{\bibinfo{volume}{26}},
  \bibinfo{pages}{2395} (\bibinfo{year}{1982}).

\bibitem[{\citenamefont{Schirmer et~al.}(1983)\citenamefont{Schirmer,
  Cederbaum, and Walter}}]{Schirmer:GF-83}
\bibinfo{author}{\bibfnamefont{J.}~\bibnamefont{Schirmer}},
  \bibinfo{author}{\bibfnamefont{L.~S.} \bibnamefont{Cederbaum}},
  \bibnamefont{and} \bibinfo{author}{\bibfnamefont{O.}~\bibnamefont{Walter}},
  \bibinfo{journal}{Phys. Rev.~A} \textbf{\bibinfo{volume}{28}},
  \bibinfo{pages}{1237} (\bibinfo{year}{1983}).

\bibitem[{\citenamefont{Schirmer and Angonoa}(1989)}]{Schirmer:SE-89}
\bibinfo{author}{\bibfnamefont{J.}~\bibnamefont{Schirmer}} \bibnamefont{and}
  \bibinfo{author}{\bibfnamefont{G.}~\bibnamefont{Angonoa}},
  \bibinfo{journal}{J. Chem. Phys.} \textbf{\bibinfo{volume}{91}},
  \bibinfo{pages}{1754} (\bibinfo{year}{1989}).

\bibitem[{\citenamefont{Weikert et~al.}(1996)\citenamefont{Weikert, Meyer,
  Cederbaum, and Tarantelli}}]{Weikert:BL-96}
\bibinfo{author}{\bibfnamefont{H.-G.} \bibnamefont{Weikert}},
  \bibinfo{author}{\bibfnamefont{H.-D.} \bibnamefont{Meyer}},
  \bibinfo{author}{\bibfnamefont{L.~S.} \bibnamefont{Cederbaum}},
  \bibnamefont{and}
  \bibinfo{author}{\bibfnamefont{F.}~\bibnamefont{Tarantelli}},
  \bibinfo{journal}{J. Chem. Phys.} \textbf{\bibinfo{volume}{104}},
  \bibinfo{pages}{7122} (\bibinfo{year}{1996}).

\bibitem[{\citenamefont{Angonoa et~al.}(1987)\citenamefont{Angonoa, Walter, and
  Schirmer}}]{Angonoa:KS-87}
\bibinfo{author}{\bibfnamefont{G.}~\bibnamefont{Angonoa}},
  \bibinfo{author}{\bibfnamefont{O.}~\bibnamefont{Walter}}, \bibnamefont{and}
  \bibinfo{author}{\bibfnamefont{J.}~\bibnamefont{Schirmer}},
  \bibinfo{journal}{J. Chem. Phys.} \textbf{\bibinfo{volume}{87}},
  \bibinfo{pages}{6789} (\bibinfo{year}{1987}).

\bibitem[{\citenamefont{Schirmer and Thiel}(2001)}]{Schirmer:ISR-01}
\bibinfo{author}{\bibfnamefont{J.}~\bibnamefont{Schirmer}} \bibnamefont{and}
  \bibinfo{author}{\bibfnamefont{A.}~\bibnamefont{Thiel}}, \bibinfo{journal}{J.
  Chem. Phys.} \textbf{\bibinfo{volume}{115}}, \bibinfo{pages}{10621}
  (\bibinfo{year}{2001}).

\bibitem[{\citenamefont{Cederbaum et~al.}(1980)\citenamefont{Cederbaum, Domcke,
  and Schirmer}}]{Cederbaum:CH-80}
\bibinfo{author}{\bibfnamefont{L.~S.} \bibnamefont{Cederbaum}},
  \bibinfo{author}{\bibfnamefont{W.}~\bibnamefont{Domcke}}, \bibnamefont{and}
  \bibinfo{author}{\bibfnamefont{J.}~\bibnamefont{Schirmer}},
  \bibinfo{journal}{Phys. Rev.~A} \textbf{\bibinfo{volume}{22}},
  \bibinfo{pages}{206} (\bibinfo{year}{1980}).

\bibitem[{\citenamefont{Tarantelli}()}]{Tarantelli:BP-03}
\bibinfo{author}{\bibfnamefont{F.}~\bibnamefont{Tarantelli}}
  \bibinfo{note}{(private communication)}.

\bibitem[{gam()}]{gamess-uk}
\bibinfo{note}{\textsc{Gamess-UK} is a package of \emph{ab initio} programs 
  written by M. F. Guest, J. H. van Lenthe, J. Kendrick, K. Schoffel, and P. 
  Sherwood, with contributions from R. D. Amos, R. J. Buenker, H. J. J. van Dam, 
  M. Dupuis, N. C. Handy, I. H. Hillier, P. J. Knowles, V. Bonacic-Koutecky, W. 
  von Niessen, R. J. Harrison, A. P. Rendell, V. R. Saunders, A. J. Stone, D. J. 
  Tozer, and A. H. de Vries. The package is derived from the original 
  \textsc{gamess} code due to M. Dupuis, D. Spangler and J. Wendoloski, NRCC 
  Software Catalog, Vol.~1, Program No.~QG01 (\textsc{gamess}), 1980.}

\bibitem[{\citenamefont{Buth}(2002)}]{Buth:NH-02}
\bibinfo{author}{\bibfnamefont{C.}~\bibnamefont{Buth}},
  \bibinfo{type}{Diplomarbeit}, \bibinfo{school}{Ruprecht-Karls Universit\"at
  Heidelberg}, \bibinfo{address}{Theoretische Chemie, Physikalisch-Chemisches
  Institut, Im Neuenheimer Feld~229, 69120~Heidelberg, Germany}
  (\bibinfo{year}{2002}),
  \eprint{\href{http://www.ub.uni-heidelberg.de/archiv/3004}
  {www.ub.uni-heidelberg.de/archiv/3004}}.

\bibitem[{\citenamefont{Godbout et~al.}(1992)\citenamefont{Godbout, Salahub,
  Andzelm, and Wimmer}}]{Godbout:GT-92}
\bibinfo{author}{\bibfnamefont{N.}~\bibnamefont{Godbout}},
  \bibinfo{author}{\bibfnamefont{D.~R.} \bibnamefont{Salahub}},
  \bibinfo{author}{\bibfnamefont{J.}~\bibnamefont{Andzelm}}, \bibnamefont{and}
  \bibinfo{author}{\bibfnamefont{E.}~\bibnamefont{Wimmer}},
  \bibinfo{journal}{Can. J. Chem.} \textbf{\bibinfo{volume}{70}},
  \bibinfo{pages}{560} (\bibinfo{year}{1992}).

\bibitem[{bas()}]{basislib}
\bibinfo{note}{Basis sets were obtained from the \emph{Extensible Computational
  Chemistry Environment Basis Set Database}, Version 5/22/02, as developed and
  distributed by the Molecular Science Computing Facility, Environmental and
  Molecular Sciences Laboratory which is part of the Pacific Northwest
  Laboratory, P.O. Box 999, Richland, Washington~99352, USA, and funded by the
  U.S. Department of Energy. The Pacific Northwest Laboratory is a
  multi-program laboratory operated by Battelle Memorial Institute for the U.S.
  Department of Energy under contract~DE-AC06-76RLO~1830. Contact David Feller
  or Karen Schuchardt for further information.}

\bibitem[{\citenamefont{Pyykk\"o}(1988)}]{Pyykko:RE-88}
\bibinfo{author}{\bibfnamefont{P.}~\bibnamefont{Pyykk\"o}},
  \bibinfo{journal}{Chem. Rev.} \textbf{\bibinfo{volume}{88}},
  \bibinfo{pages}{563} (\bibinfo{year}{1988}).

\bibitem[{\citenamefont{Balasubramanian}(1997)}]{Balasubramanian:RE-97}
\bibinfo{author}{\bibfnamefont{K.}~\bibnamefont{Balasubramanian}},
  \emph{\bibinfo{title}{Relativistic Effects in Chemistry: Theory and
  Techniques}} (\bibinfo{publisher}{John Wiley \& Sons}, \bibinfo{address}{New
  York}, \bibinfo{year}{1997}), ISBN \bibinfo{isbn}{0-471-30400-X}.

\bibitem[{\citenamefont{Froese-Fischer}(1978)}]{Froese:HF-78}
\bibinfo{author}{\bibfnamefont{C.}~\bibnamefont{Froese-Fischer}},
  \bibinfo{journal}{Comput. Phys. Commun.} \textbf{\bibinfo{volume}{14}},
  \bibinfo{pages}{145} (\bibinfo{year}{1978}).

\bibitem[{\citenamefont{Dyall et~al.}(1989)\citenamefont{Dyall, Grant, Johnson,
  Parpia, and Plummer}}]{Dyall:GR-89}
\bibinfo{author}{\bibfnamefont{K.~G.} \bibnamefont{Dyall}},
  \bibinfo{author}{\bibfnamefont{I.~P.} \bibnamefont{Grant}},
  \bibinfo{author}{\bibfnamefont{T.}~\bibnamefont{Johnson}, \bibfnamefont{C.}},
  \bibinfo{author}{\bibfnamefont{F.~A.} \bibnamefont{Parpia}},
  \bibnamefont{and} \bibinfo{author}{\bibfnamefont{E.~P.}
  \bibnamefont{Plummer}}, \bibinfo{journal}{Comput. Phys. Commun.}
  \textbf{\bibinfo{volume}{55}}, \bibinfo{pages}{425} (\bibinfo{year}{1989}).

\bibitem[{\citenamefont{Parpia et~al.}(1996)\citenamefont{Parpia,
  Froese-Fischer, and Grant}}]{Parpia:GR-96}
\bibinfo{author}{\bibfnamefont{F.~A.} \bibnamefont{Parpia}},
  \bibinfo{author}{\bibfnamefont{C.}~\bibnamefont{Froese-Fischer}},
  \bibnamefont{and} \bibinfo{author}{\bibfnamefont{I.~P.} \bibnamefont{Grant}},
  \bibinfo{journal}{Comput. Phys. Commun.} \textbf{\bibinfo{volume}{94}},
  \bibinfo{pages}{249} (\bibinfo{year}{1996}).

\bibitem[{\citenamefont{Koopmans}(1933)}]{Koopmans:-33}
\bibinfo{author}{\bibfnamefont{T.}~\bibnamefont{Koopmans}},
  \bibinfo{journal}{Physica} \textbf{\bibinfo{volume}{1}}, \bibinfo{pages}{104}
  (\bibinfo{year}{1933}).

\bibitem[{\citenamefont{Mulliken}(1955)}]{Mulliken:-55}
\bibinfo{author}{\bibfnamefont{R.~S.} \bibnamefont{Mulliken}},
  \bibinfo{journal}{J. Chem. Phys.} \textbf{\bibinfo{volume}{23}},
  \bibinfo{pages}{1833} (\bibinfo{year}{1955}).

\bibitem[{\citenamefont{Sakurai}(1994)}]{Sakurai:MQM-94}
\bibinfo{author}{\bibfnamefont{J.~J.} \bibnamefont{Sakurai}},
  \emph{\bibinfo{title}{Modern Quantum Mechanics}}
  (\bibinfo{publisher}{Addison-Wesley}, \bibinfo{address}{Reading
  (Massachusetts)}, \bibinfo{year}{1994}), \bibinfo{edition}{2nd} ed., ISBN
  \bibinfo{isbn}{0-201-53929-2}.

\bibitem[{\citenamefont{Cederbaum et~al.}(1986)\citenamefont{Cederbaum, Domcke,
  Schirmer, and von Niessen}}]{Cederbaum:CE-86}
\bibinfo{author}{\bibfnamefont{L.~S.} \bibnamefont{Cederbaum}},
  \bibinfo{author}{\bibfnamefont{W.}~\bibnamefont{Domcke}},
  \bibinfo{author}{\bibfnamefont{J.}~\bibnamefont{Schirmer}}, \bibnamefont{and}
  \bibinfo{author}{\bibfnamefont{W.}~\bibnamefont{von Niessen}}, in
  \emph{\bibinfo{booktitle}{Adv. Chem. Phys.}}, edited by
  \bibinfo{editor}{\bibfnamefont{I.}~\bibnamefont{Prigogine}} \bibnamefont{and}
  \bibinfo{editor}{\bibfnamefont{S.~A.} \bibnamefont{Rice}}
  (\bibinfo{publisher}{John Wiley \& Sons}, \bibinfo{year}{1986}),
  vol.~\bibinfo{volume}{65}, pp. \bibinfo{pages}{115--159}.

\bibitem[{\citenamefont{Zobeley et~al.}(1998)\citenamefont{Zobeley, Cederbaum,
  and Tarantelli}}]{Zobeley:HE-98}
\bibinfo{author}{\bibfnamefont{J.}~\bibnamefont{Zobeley}},
  \bibinfo{author}{\bibfnamefont{L.~S.} \bibnamefont{Cederbaum}},
  \bibnamefont{and}
  \bibinfo{author}{\bibfnamefont{F.}~\bibnamefont{Tarantelli}},
  \bibinfo{journal}{J. Chem. Phys.} \textbf{\bibinfo{volume}{108}},
  \bibinfo{pages}{9737} (\bibinfo{year}{1998}).

\bibitem[{\citenamefont{Santra et~al.}(2001)\citenamefont{Santra, Zobeley, and
  Cederbaum}}]{Santra:ED-01}
\bibinfo{author}{\bibfnamefont{R.}~\bibnamefont{Santra}},
  \bibinfo{author}{\bibfnamefont{J.}~\bibnamefont{Zobeley}}, \bibnamefont{and}
  \bibinfo{author}{\bibfnamefont{L.~S.} \bibnamefont{Cederbaum}},
  \bibinfo{journal}{Phys. Rev.~B} \textbf{\bibinfo{volume}{64}},
  \bibinfo{pages}{245104} (\bibinfo{year}{2001}).

\bibitem[{\citenamefont{Buth et~al.}(2003)\citenamefont{Buth, Santra, and
  Cederbaum}}]{Buth:-03}
\bibinfo{author}{\bibfnamefont{C.}~\bibnamefont{Buth}},
  \bibinfo{author}{\bibfnamefont{R.}~\bibnamefont{Santra}}, \bibnamefont{and}
  \bibinfo{author}{\bibfnamefont{L.~S.} \bibnamefont{Cederbaum}},
  \bibinfo{journal}{J. Chem. Phys.} \textbf{\bibinfo{volume}{119}},
  \bibinfo{pages}{10575} (\bibinfo{year}{2003}),
  \eprint{arXiv: physics/0303100}.

\bibitem[{\citenamefont{Santra and Cederbaum}(2002)}]{Santra:NH-02}
\bibinfo{author}{\bibfnamefont{R.}~\bibnamefont{Santra}} \bibnamefont{and}
  \bibinfo{author}{\bibfnamefont{L.~S.} \bibnamefont{Cederbaum}},
  \bibinfo{journal}{Phys. Rep.} \textbf{\bibinfo{volume}{368}},
  \bibinfo{pages}{1} (\bibinfo{year}{2002}).

\bibitem[{\citenamefont{Coville and Thomas}(1991)}]{Coville:ME-91}
\bibinfo{author}{\bibfnamefont{M.}~\bibnamefont{Coville}} \bibnamefont{and}
  \bibinfo{author}{\bibfnamefont{T.~D.} \bibnamefont{Thomas}},
  \bibinfo{journal}{Phys. Rev.~A} \textbf{\bibinfo{volume}{43}},
  \bibinfo{pages}{6053} (\bibinfo{year}{1991}).

\bibitem[{\citenamefont{Hartmann}(1988)}]{Hartmann:MC-88}
\bibinfo{author}{\bibfnamefont{E.}~\bibnamefont{Hartmann}},
  \bibinfo{journal}{J. Phys.~B} \textbf{\bibinfo{volume}{21}},
  \bibinfo{pages}{1173} (\bibinfo{year}{1988}).

\bibitem[{\citenamefont{Carroll et~al.}(2002)\citenamefont{Carroll, B\o{}rve,
  S\ae{}thre, Bozek, Kukk, Hahne, and Thomas}}]{Carroll:PE-02}
\bibinfo{author}{\bibfnamefont{T.~X.} \bibnamefont{Carroll}},
  \bibinfo{author}{\bibfnamefont{K.~J.} \bibnamefont{B\o{}rve}},
  \bibinfo{author}{\bibfnamefont{L.~J.} \bibnamefont{S\ae{}thre}},
  \bibinfo{author}{\bibfnamefont{J.~D.} \bibnamefont{Bozek}},
  \bibinfo{author}{\bibfnamefont{E.}~\bibnamefont{Kukk}},
  \bibinfo{author}{\bibfnamefont{J.~A.} \bibnamefont{Hahne}}, \bibnamefont{and}
  \bibinfo{author}{\bibfnamefont{T.~D.} \bibnamefont{Thomas}},
  \bibinfo{journal}{J. Chem. Phys.} \textbf{\bibinfo{volume}{116}},
  \bibinfo{pages}{10221} (\bibinfo{year}{2002}).

\bibitem[{\citenamefont{Santra et~al.}(2000)\citenamefont{Santra, Zobeley,
  Cederbaum, and Moiseyev}}]{Santra:ICD-00}
\bibinfo{author}{\bibfnamefont{R.}~\bibnamefont{Santra}},
  \bibinfo{author}{\bibfnamefont{J.}~\bibnamefont{Zobeley}},
  \bibinfo{author}{\bibfnamefont{L.~S.} \bibnamefont{Cederbaum}},
  \bibnamefont{and} \bibinfo{author}{\bibfnamefont{N.}~\bibnamefont{Moiseyev}},
  \bibinfo{journal}{Phys. Rev. Lett.} \textbf{\bibinfo{volume}{85}},
  \bibinfo{pages}{4490} (\bibinfo{year}{2000}).

\bibitem[{\citenamefont{Marburger et~al.}(2003)\citenamefont{Marburger,
  Kugeler, Hergenhahn, and M\"oller}}]{Marburger:EE-03}
\bibinfo{author}{\bibfnamefont{S.}~\bibnamefont{Marburger}},
  \bibinfo{author}{\bibfnamefont{O.}~\bibnamefont{Kugeler}},
  \bibinfo{author}{\bibfnamefont{U.}~\bibnamefont{Hergenhahn}},
  \bibnamefont{and} \bibinfo{author}{\bibfnamefont{T.}~\bibnamefont{M\"oller}},
  \bibinfo{journal}{Phys. Rev. Lett.} \textbf{\bibinfo{volume}{90}},
  \bibinfo{pages}{203401} (\bibinfo{year}{2003}).

\bibitem[{\citenamefont{Zobeley et~al.}(2001)\citenamefont{Zobeley, Santra, and
  Cederbaum}}]{Zobeley:ED-01}
\bibinfo{author}{\bibfnamefont{J.}~\bibnamefont{Zobeley}},
  \bibinfo{author}{\bibfnamefont{R.}~\bibnamefont{Santra}}, \bibnamefont{and}
  \bibinfo{author}{\bibfnamefont{L.~S.} \bibnamefont{Cederbaum}},
  \bibinfo{journal}{J. Chem. Phys.} \textbf{\bibinfo{volume}{115}},
  \bibinfo{pages}{5076} (\bibinfo{year}{2001}).
\end{thebibliography}
\end{document}